\documentstyle[12pt]{article}
\renewcommand{\theequation}{\thesection.\arabic{equation}}
\newcommand{\beq}{\begin{equation}}
\newcommand{\eeq}{\end{equation}}
\newcommand{\bea}{\begin{eqnarray}}
\newcommand{\eea}{\end{eqnarray}}

\begin{document}
\setcounter{page}{0}
\topmargin 0pt
\oddsidemargin 5mm
\renewcommand{\thefootnote}{\fnsymbol{footnote}}
\newpage
\setcounter{page}{0}
\begin{titlepage}

\begin{flushright}
QMW-PH-99-01\\
%OU-TP-97-11P\\
{\bf hep-th/9901131}\\
 {\it January 1999}
\end{flushright}
\vspace{0.5cm}
\begin{center}
{\Large {\bf On zigzag-invariant strings}} \\
\vspace{1.8cm}
\vspace{0.5cm}
{Ian I. Kogan$~^{1,}$\footnote{e-mail:
i.kogan1@physics.oxford.ac.uk} and  Oleg A.
Soloviev$~^{2,}$
\footnote{e-mail: O.A.Soloviev@QMW.AC.UK}} \\
\vspace{0.5cm}
{\em$~^1$Department of Physics, University of Oxford\\
1 Keble Road, Oxford, OX1 3NP, United Kingdom}\\
{\em$~^2$ Physics Department, Queen Mary and Westfield College, \\
Mile End Road, London E1 4NS, United Kingdom}\\
\vspace{0.5cm}
\renewcommand{\thefootnote}{\arabic{footnote}}
\setcounter{footnote}{0}
\begin{abstract}
{We propose a world-sheet realization of the zigzag-invariant 
strings as a perturbed WZNW model at large negative level. 
We argue that the gravitational dressing produces additional 
fixed points of the dressed renormalization beta function. 
One of these new critical points can be interpreted as a zigzag-invariant 
string model.
}
\end{abstract}
\vspace{0.5cm}
%\centerline{July 1996}
 \end{center}
\end{titlepage}
\newpage

\renewcommand{\theequation}{\arabic{equation}}
\setcounter{equation}{0}

A  world-sheet description of the QCD string has not as yet been discovered. At the same time, %%@
the general feeling is that we are getting very close to unveiling the one of the most %%@
intriguing enigma of theoretical physics. Recently, a major breakthrough has been achieved %%@
\cite{Polyakov},\cite{Maldacena},\cite{Gubser}. It has been shown that a Yang-Mills theory in %%@
$D$-dimensions can be obtained from a (super)gravity defined in a $D+1$-dimensional space-time. %%@
The open problem is to understand the concrete world-sheet realisation of these ideas.

Here we would like to attempt a formulation of a two-dimensional model which may have properties %%@
similar to the ones required for the QCD string. According to Polyakov, the latter has to be %%@
described by the following world-sheet action \cite{Polyakov}
\begin{equation}
S_P=\int d^2\xi\left[(\partial\phi)^2~+~a^2(\phi)(\partial x)^2~+~
\Phi(\phi)R^{(2)}\sqrt g\right],\label{Paction}\end{equation}
where we omit possible antisymmetric fields, like $B_{\mu\nu}$ or Ramond-Ramond fields in the %%@
supersymmetric case. Here $\phi$ is the Liouville field of 2D gravity, $x^\mu$ ($\mu$ runs from %%@
1 to $D$) are coordinates of the confining string, $\Phi$ is the dilaton field, $R^{(2)}$ is the %%@
curvature of the world-sheet and $a(\phi)$ is the running string tension. The zigzag symmetry %%@
requires the existence of a certain value of the Liouville field $\phi^{*}$ such that %%@
\cite{Polyakov}
\begin{equation}
a(\phi^{*})=0.\label{zigzag}\end{equation}

A concrete world-sheet realisation of Polyakov's ansatz appears to be fairly intricate. Our %%@
first step is to consider a certain non-conformal model with the running coupling constant in %%@
front of the kinetic term. Such a theory has been discussed in \cite{Soloviev1}. It is a (non-%%@
unitary) WZNW model perturbed by its kinetic term. The corresponding action is written as %%@
follows
\begin{equation}
S(\epsilon)=S_{WZNW}(G,k)~-~\epsilon\int d^2z\;O(z,\bar z).\label{perturbation}\end{equation}
Here $\epsilon$ is a small constant, $S_{WZNW}(G,k)$ is the WZNW model on the group manifold $G$ %%@
at level $k$ and
\begin{equation}
O(z,\bar z)={1\over c_V(G)}J^a\bar J^b\phi^{ab},\label{O}\end{equation}
where
\begin{equation}
J\equiv J^at^a={-k\over2}\partial gg^{-1},~
\bar J\equiv \bar J^at^a={-k\over2}g^{-1}\bar\partial g,~
\phi^{ab}=\mbox{Tr}(g^{-1}t^agt^b),~c_V(G)={-f^{ac}_df^{bd}_c\eta_{ab}\over\dim G}.
\label{definitions}\end{equation}
The operator $O$ has the following conformal dimension
\begin{equation}
\Delta_O=1~+~{c_V(G)\over k+c_V(G)}.\label{dimension}\end{equation}
Thus, for $k<-2c_V(G)$, $O$ is a relevant operator with positive conformal dimension. Another %%@
important property of $O$ is that it satisfies the following OPE
\begin{equation}
O\cdot O=[{\bf 1}]~+~[O]~+~...,\label{OPE}\end{equation}
where the square brackets denote the contribution of $O$ and the identity operator and their %%@
descendants, whereas dots stand for operators with conformal dimensions greater than one. The %%@
given OPE implies that the perturbed CFT is renormalizable.

Since the critical point $\epsilon_0=0$ is unstable, the theory (\ref{perturbation}) will flow %%@
to some IR conformal point or to a massive phase. The corresponding renormalization group beta %%@
function is given as follows
\begin{equation}
\beta\equiv{d\epsilon\over d\ln\Lambda}=(2-2\Delta_O)\epsilon~-~\pi\epsilon^2~+~...,
\label{beta}\end{equation}
where $\Lambda$ is the ultraviolet cutoff. In the limit $k\to-\infty$, eq.(\ref{beta}) has a %%@
non-trivial fixed point
\begin{equation}
\epsilon^{*}=-{2c_V(G)\over\pi k},\label{FP}\end{equation}
at which the perturbed non-unitary WZNW model becomes a unitary WZNW model with positive level %%@
$l\to|k|$.

Along the flow from $\epsilon_0$ to $\epsilon^{*}$, the parameter $\epsilon$ passes the middle %%@
point
\begin{equation}
\epsilon_{WZ}=-{c_V(G)\over\pi k},\label{WZ}\end{equation}
at which the deformation kills the sigma-model kinetic term and the resulting theory has only %%@
the Wess-Zumino term, i.e.
\begin{equation}
S(\epsilon_{WZ})\sim\Gamma.\label{Gamma}\end{equation}
This point does not appear to be critical, however, the properties of the theory at this point %%@
are quite puzzling \cite{Stern} which make very difficult its quantum interpretation. 

In spite of certain pathology, the described perturbed WZNW model seems to mimic well the %%@
behaviour of Polaykov's confining string. In order to make the link between the perturbed WZNW %%@
model and Polyakov's ansatz more precise, we have to couple the former to 2D gravity. This gives %%@
rise to the so-called gravitational dressed beta function $\bar\beta$ %%@
\cite{Klebanov},\cite{Schmidhuber},\cite{Dorn}. Usually, the latter can be written as an %%@
expansion in the perturbation parameter but only with new coefficients:
\begin{equation}
\bar\beta=(2-2\Delta)\epsilon~-~\pi\bar C\epsilon^2~+~...,\label{barbeta}\end{equation}
where $\Delta$ and $\bar C$ are the KPZ gravitational scaling dimension \cite{KPZ} and the %%@
coefficient of the dressed three-point function respectively,
\begin{equation}
\Delta-\Delta_O={\Delta(\Delta-1)\over\kappa+2},~~~
\bar C={2\over\alpha(Q+2\alpha_1)}.\label{KPZ}\end{equation}

Here
\begin{eqnarray}
\kappa+2={1\over12}\left(c-13+\sqrt{(c-1)(c-25)}\right),~~~Q=\sqrt{|c-25|\over3},\nonumber\\ & & %%@
\\
\alpha=-{Q\over2}~+~\sqrt{{Q^2\over4}+2},~~~\alpha_1=-{Q\over2}~+~\sqrt{{Q^2\over4}-(2\Delta_O-%%@
2)}.\nonumber\label{dressing}\end{eqnarray}
We also assume that $c>25$.

For large $c$, the dressed beta function $\bar\beta$ has exactly the same fixed points as the %%@
original beta function $\beta$. This can be seen either from direct calculation or from the %%@
following formula \cite{Tseytlin}
\begin{equation}
\dot{\bar\beta}~+~Q\bar\beta~+~{\cal O}(\bar\beta^2)=\beta.\label{main}\end{equation}
However, when $c$ approaches 25, the expansion (\ref{barbeta}) is no longer valid as the %%@
corresponding coefficients become singular in the limit $k\to-\infty$. For example, one can show %%@
that for $c=25$, the dressed beta function of a marginal ($\Delta=1$) operator behaves as %%@
follows
\begin{equation}
\bar\beta(\Delta=1)\sim\epsilon^{3/2}~+~...\label{3/2}\end{equation}
Thus, around 25, the expansion of the gravitational dressed beta function in $\epsilon$ may %%@
completely change its form, e.g. instead of integer powers of $\epsilon$, there might be integer %%@
and half-integer powers of $\epsilon$. This may lead to new critical points of the dressed beta %%@
function compared with the original beta function, that is zeros of $\bar\beta$ will not %%@
necessarily be zeros of $\beta$. This situation is realised in the case under consideration.

In order to see that there are new zeros, we shall use two more very important equations %%@
\cite{Tseytlin}
\begin{equation}
Q^2={c(\Lambda)-25\over3}~+~{1\over4}\bar\beta^2,~~~
\dot Q=-{1\over4}\bar\beta^2,\label{two}\end{equation}
where $c(\Lambda)$ is Zamolodchikov's c-function which changes along the flow (as a function of %%@
the cut-off $\Lambda$).
In models with gravity, the c-function does not necessarily decreases, however the second %%@
equation in (\ref{two}) tells us that the function $Q(\Lambda)$ does decrease along the flow. 

Let us now consider the situation when $c$ is slightly greater than 25. This can be realised, if %%@
we take a perturbed WZNW model and add to it a CFT whose Virasoro central charge is equal to %%@
$25-\dim G$. Then the total Virasoro central charge is given as follows
\begin{equation}
c=25~+~{c_V(G)\dim G\over|k|}~+~...>25.\label{>25}\end{equation}
Without gravity, the flow would take us to the IR conformal point with $c<25$. These two fixed %%@
points are symmetrically located with respect to the middle point at which $c=25$. However, when %%@
the gravity is turned on, the flow cannot end under the $c=25$ barrier \cite{Tseytlin}, although %%@
it can cross it. Without gravity, the middle point $\epsilon_{WZ}$ corresponds to 
\begin{equation}
c(\epsilon_{WZ})=25,\label{QWZ}\end{equation}
whereas $\epsilon^{*}$ gives rise to the negative $Q$.
When we turn on 2D gravity, we can no longer reliably use the expansion (\ref{barbeta}) around %%@
$c=25$. However, the information about the fixed points far away from 25 allows us to establish %%@
that $\epsilon_{WZ}$ must remain in the middle also when gravity is turned on. Thus, in the %%@
presence of gravity eq.(\ref{QWZ}) still has to hold. Taking into account  equations %%@
(\ref{two}), we find
\begin{equation}
Q(\epsilon_{WZ})=0,~~~\bar\beta|_{\epsilon_{WZ}}=0.\label{new}\end{equation}
All in all, we arrive at the conclusion that the middle point (\ref{WZ}) is a conformal point of %%@
the dressed beta function. The appearance of this additional conformal point prevents the flow %%@
from crossing the barrier. This point is very nontrivial. In particular, we expect that %%@
eq.(\ref{main}) has to break down at the given point. One reason for it is that this equation %%@
was derived within the $\alpha'$-expansion \cite{Tseytlin}. However, at this point the string %%@
tension vanishes, that is $\alpha'$ becomes very large. Unfortunately, it appears to be very %%@
difficult to carry out any qualitative analysis around $\epsilon_{WZ}$. The other two equations %%@
in (\ref{two}) seem to be still valid, since they are not directly affected by the running %%@
string tension. 

Note that our analysis works for any group $G$. Perhaps $G$ has to be chosen noncompact because %%@
we are obliged to deal with negative $k$. Thus, we can lift off the restriction on the group %%@
manifold obtained in our previous paper \cite{Kogan}.

\vskip0.5cm
\noindent
{\large \bf Acknowledgements}

\smallskip
\noindent
We like to thank the orginizers of Buckow-98 for a very nice conference and also the orginizers %%@
of the conference "TRENDS in MATHEMATICAL PHYSICS" in the
University of Tennesse at Knoxville, October 98 for the opportunity to present these results.

%%%%%%%%%%%%%%%%%%%%%%
%%%%%%%%%%%%%%%%%%%%%%

\end{document}